\documentclass[aps,pra,showpacs,twocolumn,superscriptaddress]{revtex4}

\usepackage{amsmath,amssymb,amsthm}
\usepackage{graphicx}
\usepackage{color}
\usepackage{tabularx}

\newcommand{\myup}[2]{#1^{{\color{white}\dagger \hspace{-0.8ex}}#2}}

\begin{document}
\title{Measuring arbitrary-order coherences: Tomography of single-mode multiphoton polarization-entangled states}

\author{U. Schilling}
\affiliation{Institut f\"{u}r Optik, Information und Photonik, Universit\"{a}t Erlangen-N\"{u}rnberg, 91058 Erlangen, Germany}

\author{J. von Zanthier}
\affiliation{Institut f\"{u}r Optik, Information und Photonik, Universit\"{a}t Erlangen-N\"{u}rnberg, 91058 Erlangen, Germany}

\author{G. S. Agarwal}
\affiliation{Department of Physics, Oklahoma State University, Stillwater, Oklahoma 74078-3072, USA}

\begin{abstract}
A scheme is discussed for measuring $N$th-order coherences of two orthogonally polarized light fields in a single spatial mode at very limited experimental cost. To implement the scheme, the only measurements needed are the $N$th-order intensity moments after the light beam has passed through two quarter-wave plates, one half-wave plate, and a polarizing beam splitter for specific settings of the wave plates. It is shown that this method can be applied for arbitrarily large $N$. A set of explicit values is given for the settings of the wave plates, constituting an optimal measurement of the $N$th-order coherences for any $N$. For Fock states the method introduced here corresponds to a full state tomography. Applications of the scheme to systems other than polarization optics are discussed.
\end{abstract}

\maketitle

\section{Introduction}
In recent years, interest in entangled states has grown significantly. One of the most successful experimental implementations has been achieved by entangling the polarization degrees of freedom of two photons in spontaneous parametric down conversion~\cite{Kwiat:1995}. Since then, a lot of effort has been devoted to producing entangled states with an ever higher photon number~\cite{Mitchell:2004,Walther:2004,Lu:2007,Wieczorek:2009,Eisenberg:2004}. For the most part, entangled photon states are generated by postselection in such a way that every photon is found to have occupied an individual spatial mode~\cite{Weinfurter:2007}. In that kind of setup, a quantum state tomography is usually conducted with the help of quarter-wave plates and half-wave plates which operate on every output port separately in such a way that each mode is analyzed along different orthogonal bases. The theory behind this method is intuitive and has been described exhaustively~\cite{Kwiat:2005}. However, there exist experiments that generate polarization-entangled states of higher photon numbers in a single spatial mode~\cite{Tsegaye:2000,Steinberg:2009}. Here, it is possible to split the beam into as many spatial modes as there are photons to conduct a full state tomography as in \cite{Kwiat:2005}, but for higher photon numbers, this approach is experimentally very costly and inefficient.

In this paper, we discuss a method that does not require the beam to be separated in different spatial modes. It is based on a very general theorem, formulated by Mukunda and Jordan, which states that it is always possible to calculate the coherences of a photon field from photon correlation measurements in several different bases~\cite{Mukunda:1966}. This theorem has been used in proposals to measure all second-order coherences of a field, corresponding to the variances of the so-called quantum Stokes parameters~\cite{Agarwal:2003,Rivas:2008}. Here we show that it is possible to measure \emph{all} $N$th-order coherences in a light beam consisting of two polarization modes with an arbitrary and unknown amount of photons in each mode. For this measurement we only require two quarter-wave plates, one half-wave plate, and a polarizing beam splitter, all acting on the same single spatial mode. The $N$th-order intensity moment of that mode is measured after passage of the photons through the mentioned optical elements~\cite{Vogel:2006}. If the incoming field is in a Fock state of $N$ photons, this procedure corresponds to a full state tomography of that state~\cite{Steinberg:2008}. We note that the scheme is not limited to polarization optics, but may also be applied to other two-mode systems  where the necessary operations can be implemented, for example photons in two different Laguerre-Gaussian modes~\cite{Lassen:2009,Hsu:2009}.

\section{The Basis of the Method}
\label{basics}
An often-used description for the polarization state of light are the Stokes parameters, which describe the state of a polarized light beam as a point on the Poincar\'e sphere. For a classical coherent beam, the parameters are defined by:
\begin{align}
\begin{array}{ll}
S_0 = |\alpha_1|^2 + |\alpha_2|^2, \hspace{2em}& S_1 = |\alpha_1|^2 - |\alpha_2|^2,\\
S_2 = \alpha_1^* \alpha_2 + \alpha_1 \alpha_2^*,& S_3 = -i(\alpha_1^* \alpha_2 - \alpha_1 \alpha_2^*),
\end{array}
\end{align}
where $\alpha_1$ and $\alpha_2$ represent the amplitude of the beam in two orthogonal linear polarizations. The quantum Stokes parameters are derived from the classical description by replacing the amplitudes and their complex conjugates with the annihilation and creation operators of the electric field components. This leads to the following definition for the quantum Stokes parameters:
\begin{align}
\begin{array}{ll}
S_0 = a_1^\dagger a_1 + a_2^\dagger a_2,\hspace{2em}&
S_1 = a_1^\dagger a_1 - a_2^\dagger a_2,\\
S_2 = a_1^\dagger a_2 + a_2^\dagger a_1,&
S_3 = -i (a_1^\dagger a_2 - a_2^\dagger a_1).
\end{array}
\end{align}
As usual, for the electric field operators the bosonic commutation relation $[a_i,a_j^\dagger] = \delta_{ij}$ holds. The measurement of the quantum Stokes parameters involves the four averages $\langle a_1^\dagger a_1 \rangle$, $\langle a_1^\dagger a_2 \rangle$, $\langle a_2^\dagger a_1 \rangle$, and $\langle a_2^\dagger a_2 \rangle$. Schemes to measure these four quantities are standard textbook material~\cite{Born:1999}, since the procedure is equivalent to determining the polarization of a light beam and their values equal those of their classical counterparts. However, the variances of the quantum Stokes parameters, defined by the difference of two anticommutators
\begin{align}
V_{ij} = \frac{1}{2}(\langle \{ S_i,S_j\} \rangle - \{ \langle S_i \rangle, \langle S_j \rangle \})
\label{stokes_variances}
\end{align}
with $i,j = \{0,1,2,3\},$ are not equal to their classical counterparts. For example, for a coherent beam, in the classical picture all variances should vanish, whereas the analysis of the quantized fields shows that the variances actually obey an uncertainty relation and cannot all vanish at the same time~\cite{Korolkova:2002}. They can be described in terms of all nine normally ordered second-order field correlations: $\langle a_1^\dagger a_1^\dagger a_1 a_1\rangle$, $\langle a_1^\dagger a_1^\dagger a_1 a_2\rangle$, $\langle a_1^\dagger a_2^\dagger a_1 a_1\rangle$, $\langle a_1^\dagger a_2^\dagger a_1 a_2\rangle$, $\langle a_1^\dagger a_1^\dagger a_2 a_2\rangle$, $\langle a_1^\dagger a_2^\dagger a_2 a_2\rangle$, $\langle a_2^\dagger a_2^\dagger a_1 a_2\rangle$, $\langle a_2^\dagger a_2^\dagger a_1 a_1\rangle$, and $\langle a_2^\dagger a_2^\dagger a_2 a_2\rangle$. Out of those, only the three field correlations $\langle a_1^\dagger a_1^\dagger a_1 a_1\rangle$, $\langle a_1^\dagger a_2^\dagger a_1 a_2\rangle$, and $\langle a_2^\dagger a_2^\dagger a_2 a_2\rangle$, which correspond to second-order intensity moment measurements, are directly accessible in experiments.

Korolkova \emph{et al.} first proposed a way to measure the diagonal variances $V_{ii}$ by using a half-wave plate and a quarter-wave plate \cite{Korolkova:2002}. Later, in \cite{Agarwal:2003} it was shown that it is possible to measure the variances of all quantum Stokes parameters (i.e., all second-order coherences of light) by conducting measurements of intensity-intensity correlations after the light has passed two quarter-wave plates and one half-wave plate for a set of nine specific positions of the three wave plates, a setup which constitutes a universal SU(2) gadget for polarized light and implements a general rotation in SU(2) space~\cite{Simon:1990}. The knowledge of $\vec{S}$ and $\hat{V}$ already gives a good idea of the nature of the quantum state; however, the measurement of higher-order coherences may add even more information about that state. In particular, if the measured beam is in a photon-number state with $N$ photons, that is, if the photonic state is of the form
\begin{align}
\sum_{n=0}^N c_n |n\rangle_1 |N-n\rangle_2, \hspace{3em} \sum_{n=0}^{N} |c_n|^2 = 1,
\end{align}
the density matrix of that state has a size of $(N+1) \times (N+1)$ and its $(N+1)^2$ elements correspond to all $N$th-order coherences. Thus, for an $N$-photon Fock state, the measurement of all $N$th-order coherences is equivalent to a full state tomography.
\begin{figure}[t]
\centering
\includegraphics[scale=0.4]{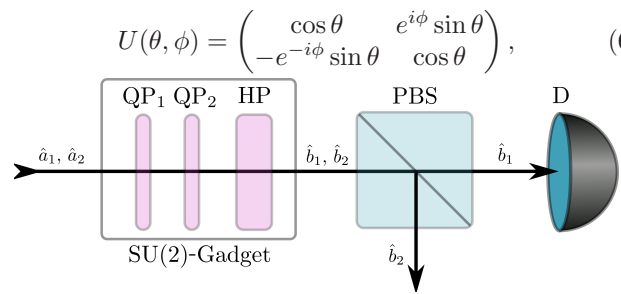}
\caption{(color online) A sketch of the setup with two quarter-wave plates (QP$_1$ and QP$_2$), one half-wave plate (HP), and a polarizing beam splitter (PBS). At the detector (D), the $N$th-order intensity moment is measured.}
\label{setup}
\end{figure}
In the following, we show that with the setup depicted in Fig.\,\ref{setup}, which is slightly modified with respect to the one discussed in \cite{Agarwal:2003}, it is possible to determine \emph{all $N$th-order coherences} for arbitrary $N$ by measuring only $N$th-order \emph{intensity moments}.

The action of the SU(2) gadget on two orthogonally polarized modes $\hat{a}_1$ and $\hat{a}_2$ can be para\-me\-trized by two angles $\theta$ and $\phi$
\begin{align}
\begin{pmatrix}b_1\\ b_2\end{pmatrix} = U(\theta,\phi) \begin{pmatrix} a_1\\ a_2\end{pmatrix}
\label{unitary_transformation}
\end{align}
with
\begin{align}
U(\theta,\phi) = \begin{pmatrix} \cos \theta & e^{i \phi}\sin \theta\\ -e^{-i\phi} \sin\theta & \cos \theta \end{pmatrix},
\end{align}
where $\theta$ and $\phi$ are abstract parameters determined by the orientation of the three wave plates. The exact functional dependence is given in Sec.\,\ref{sec_example}. Using this parametrization of the unitary transformation, we can express the most general case of measuring the $N$th-order correlation -- defined by the correlation of the $i$th intensity moment in mode $\hat{b}_1$ with the $(N-i)$th intensity moment in mode $\hat{b}_2$ -- behind the SU(2) gadget as
\begin{widetext}
\begin{multline}
\langle b_1^{\dagger i} b_2^{\dagger N-i} \myup{b_1}{i} \myup{b_2}{N-i}\rangle =
\sum_{w,y=0}^{i} \sum_{x,z=0}^{N-i} \binom{i}{w}\binom{i}{y} \binom{N-i}{x} \binom{N-i}{z}
(\cos \theta)^{2N-w-x-y-z}\\
\times (\sin \theta)^{w+x+y+z}(-1)^{x+z} e^{i\phi (x+y-w-z)} \langle a_1^{\dagger i+x-w} a_2^{\dagger N-i-x+w} \myup{a_1}{i+z-y}\myup{a_2}{N-i-z+y}\rangle.
\label{general_final_state}
\end{multline}
\end{widetext}
To solve for the $(N+1)^2$ independent real variables in the density matrix, we must perfom at least $(N+1)^2$ measurements. Hereby, we must be sure that the values of $\theta$ and $\phi$ chosen for these measurements lead to a system of independent linear equations; from Eq.\,(\ref{general_final_state}), it is not obvious that this is possible for arbitrary $N$. In the following, a set of values for $\theta$ and $\phi$ are be given for which we show that the measurement of $N$th-order intensity moments leads to a solvable system of equations. In the course of this proof, a natural recipe is developed that describes how the measurement results can be easily related to the coherences and populations of the initial state.

\section{A Set of Solutions for Eq.\,(\ref{general_final_state})}
\label{solutions}
The results of this section show that it suffices to measure photons of just one polarization, either $\hat{b}_1$ or $\hat{b}_2$, to determine all coherences. For this reason, it suffices to set up a measurement apparatus behind only one of the two output ports of the polarizing beam splitter (cf. Fig.\,\ref{setup}). In a random pick, we choose to measure the $N$th-order intensity in mode $\hat{b}_1$ (i.e. $i=N$) and can therefore drop the summation over $x$ and $z$ in Eq.\,(\ref{general_final_state}), which consequently simplifies to:
\begin{multline}
\langle b_1^{\dagger N} \myup{b_1}{N} \rangle = \sum_{w,y=0}^{N} \binom{N}{w}\binom{N}{y}
(\cos \theta)^{2N-w-y}\\ (\sin \theta)^{w+y} e^{i\phi (y-w)} \langle a_1^{\dagger N-w} a_2^{\dagger w} \myup{a_1}{N-y} \myup{a_2}{y}\rangle.
\label{only_b_1_measurement}
\end{multline}
Experimentally, $\langle b_1^{\dagger N} \myup{b_1}{N} \rangle$ corresponds to a measurement of the $N$th-order intensity moment. Note that the number of terms in Eq.\,(\ref{only_b_1_measurement}) is $(N+1)^2$ and the expectation value of each coherence and population appears exactly once. Thus, the system of linear equations generated from this equation by $(N+1)^2$ measurements of $\langle b_1^{\dagger N} \myup{b_1}{N} \rangle$ for different $\phi$ and $\theta$ has exactly one solution if and only if we can choose the values of every pair $(\phi,\theta$) such that all equations are independent. To arrive at such a choice, we first introduce new indices of summation, $\alpha$ and $\beta$, such that we can rewrite Eq.\,(\ref{only_b_1_measurement}) in a form where the phase $e^{i \beta \phi}$ factors out of one sum:
\begin{multline}
\langle b_1^{\dagger N} \myup{b_1}{N} \rangle = \sum_{\beta=-N}^N e^{i \beta \phi} \sum_{\alpha \in G_\beta} \binom{N}{\frac{\alpha+\beta}{2}} \binom{N}{\frac{\alpha-\beta}{2}}
(\cos \theta)^{2N-\alpha}\\ (\sin \theta)^\alpha \langle a_1^{\dagger N-\frac{\alpha-\beta}{2}} a_2^{\dagger \frac{\alpha-\beta}{2}} \myup{a_1}{N-\frac{\alpha+\beta}{2}} \myup{a_2}{\frac{\alpha+\beta}{2}}\rangle
\label{only_b_1_nice_form},
\end{multline}
with
\begin{align*}
\alpha = y + w \hspace{5em} \beta = y-w.
\end{align*}
and
\begin{align}
G_\beta=\{2(N-\kappa)-|\beta|\}\text{ with } \kappa\in \{0,1,\ldots,N-|\beta|\}
\label{def_of_G}
\end{align}
Equation (\ref{only_b_1_nice_form}) is the starting point for our analysis. Please note that for all $k-1$ $k$th roots of unity $r_{l}$ (except unity itself), the equation $\sum_{\kappa=0}^{k-1} (r_{l})^\kappa = 0$ holds.~\footnote{This can be seen instantaneously from the formula for the geometric series:$$\sum_{i=0}^{k-1} q^i = \frac{1-q^{k}}{1-q}$$.} This useful identity is exploited by choosing $\phi$ adequately to simplify Eq.\,(\ref{only_b_1_nice_form}) further and to introduce an inductive proof which shows that for suitable choices of $\phi$ and $\theta$, Eq.\,(\ref{general_final_state}) can be solved. However, in order to arrive at a complete solution, we must distinguish in the following between measuring coherences of an odd or an even order $N$.

\subsection{$N$ even}
If $N$ is even, we choose for $\phi$ the values $\phi_k = \frac{2\pi k}{N+1}$ with $k \in \{1,2,\ldots,N+1\}$. Consequently, $e^{\pm i\phi_k}$ corresponds to all $(N+1)$th roots of unity. For every choice of $\phi$, we perform a measurement for $N+1$ different values of $\theta$, with $\theta_j = \frac{j}{N+2}\frac{\pi}{2}$ and $j \in \{1,2,\ldots,N+1\}$, thus carrying out $(N+1)^2$ measurements and obtaining $(N+1)^2$ different equations from Eq.\,(\ref{only_b_1_nice_form}). By summing all equations of equal $\theta_j$, all terms from Eq.\,(\ref{only_b_1_nice_form}) with $\beta \neq 0$ cancel because of the mentioned property of the roots of unity, and the sum over $\beta$ contracts to $\beta = 0$. Thus, we are left with $N+1$ equations (one for every value of $j$) containing only the $N+1$ diagonal terms, each depending on a \emph{different} power of $\cos \theta_j$:
\begin{multline}
\frac{1}{N+1} x_{\theta_j} = \sum_{\alpha \in G_\beta} \binom{N}{\frac{\alpha}{2}} \binom{N}{\frac{\alpha}{2}}
(\cos \theta_j)^{2N-\alpha}\\
(\sin \theta_j)^\alpha \langle a_1^{\dagger N-\frac{\alpha}{2}} a_2^{\dagger \frac{\alpha}{2}} \myup{a_1}{N-\frac{\alpha}{2}} \myup{a_2}{\frac{\alpha}{2}}\rangle,
\label{N_even_populations}
\end{multline}
where $x_{\theta_j} = \sum_{\phi_k} x_{\theta_j}^{\phi_k}$ and $x_{\theta_j}^{\phi_k}$ is the result of the $N$ photon measurement $\langle b_1^{\dagger N} \myup{b_1}{N} \rangle$ for setting $\phi = \phi_k$ and $\theta = \theta_j$. This set of equations can now be solved for the diagonal terms.

We need not make more measurements to determine the other coherence terms. By first multiplying Eq.\,(\ref{only_b_1_nice_form}) by $e^{i\phi_k}$, and then adding all measurements for identical $\theta_j$, only terms with $\beta=-k$ and $\beta'=N-k+1$ survive and we arrive at
\begin{widetext}
\begin{multline}
\frac{e^{i\phi_k}}{N+1} x_{\theta_j} = \sum_{\alpha \in G_\beta} \binom{N}{\frac{\alpha+\beta}{2}} \binom{N}{\frac{\alpha-\beta}{2}} (\cos \theta_j)^{2N-\alpha} (\sin \theta_j)^\alpha \langle a_1^{\dagger N-\frac{\alpha-\beta}{2}} a_2^{\dagger \frac{\alpha-\beta}{2}} \myup{a_1}{N-\frac{\alpha+\beta}{2}} \myup{a_2}{\frac{\alpha+\beta}{2}}\rangle\\
+ \sum_{\alpha' \in G_{\beta'}} \binom{N}{\frac{\alpha'+\beta'}{2}} \binom{N}{\frac{\alpha'-\beta'}{2}} (\cos \theta_j)^{2N-\alpha'} (\sin \theta_j)^{\alpha'} \langle a_1^{\dagger N-\frac{\alpha'-\beta'}{2}} a_2^{\dagger \frac{\alpha'-\beta'}{2}} \myup{a_1}{N-\frac{\alpha'+\beta'}{2}} \myup{a_2}{\frac{\alpha'+\beta'}{2}}\rangle.
\label{N_even_coherences}
\end{multline}
\end{widetext}
Since $N$ is even, all $\alpha$ are odd, while all $\alpha'$ are even or vice versa [cf. Eq.\,(\ref{def_of_G})], leaving a total sum, in which each coherence term again depends on a different power of $\cos \theta_j$.~\footnote{Remember that each coherence term appears only once in Eq.\,(\ref{only_b_1_measurement}) and thus also maximally once in Eq.\,(\ref{N_even_coherences}).} Furthermore, the total number of coherence terms appearing in Eq.\,(\ref{N_even_coherences}) is given by $|G_\beta| + |G_{\beta'}|$ which is equal to $N+1$  for every choice of $k$. Thus, the system of linear equations generated from Eq.\,(\ref{N_even_coherences}) by inserting all $N+1$ values of $\theta_j$ is solvable. Furthermore, $\phi_k$ determines the coherences that appear in the system. It is enough to generate a system of linear equations for every $\phi_k$ with $k\in \{1,2,\ldots,N/2\}$ to solve for all coherences. Since the total number of measurements is equal to the total number of unknown variables, the presented set of values describes an optimal set of measurements.

\subsection{$N$ odd}
If $N$ is odd, the previously described approach does not work, since $\alpha$ and $\alpha'$ in Eq.\,(\ref{N_even_coherences}) are both even or odd. Because of this, different coherence terms will depend on the same power of $\cos \theta$ and it is consequently only possible to solve for their sum. Therefore, we modify the choice of our values of $\phi$ and $\theta$ slightly: we choose $N+2$ settings for $\phi$, with $\phi_k = \frac{2\pi k}{N+2}$, $k=\{1,2,\ldots,N+2\}$, so that $e^{i \phi_k}$ describes all $(N+2)$th roots of unity. For every $\phi_k$, we conduct measurements for $N$ different values of $\theta$, with $\theta_j = \frac{j}{N+1}\frac{\pi}{2}$ and $j \in \{1,2,\ldots,N\}$. For this set of $N(N+2)$ measurements, we proceed as in the case for even $N$: the sum of all measurements for constant $\theta$ yields
\begin{multline}
\frac{1}{N+2} x_{\theta_j} = \sum_{\alpha \in G_\beta} \binom{N}{\frac{\alpha}{2}} \binom{N}{\frac{\alpha}{2}}
(\cos \theta_j)^{2N-\alpha}\\
(\sin \theta_j)^\alpha \langle a_1^{\dagger N-\frac{\alpha}{2}} a_2^{\dagger \frac{\alpha}{2}} \myup{a_1}{N-\frac{\alpha}{2}}\myup{a_2}{\frac{\alpha}{2}}\rangle,
\label{N_odd_populations}
\end{multline}
which is identical to Eq.\,(\ref{N_even_populations}). However, we have only $N$ equations to solve for $N+1$ terms, so we must conduct one more measurement [e.g. for $(\theta,\phi) = (0,0)$] to solve for all unknowns. At this point, the total number of measurements is again $N(N+2)+1 = (N+1)^2$ and, thus, also optimal. In the following, we need not make more measurements but can directly solve for the remaining unknown variables. In a first step, we multiply all equations by $e^{i \phi_1}$ before summation and arrive at
\begin{multline}
\frac{e^{i \phi_1}}{N+2} x_{\theta_j} = \sum_{\alpha \in G_\beta} \binom{N}{\frac{\alpha-1}{2}} \binom{N}{\frac{\alpha+1}{2}} (\cos \theta_j)^{2N-\alpha}\\
(\sin \theta_j)^\alpha \langle a_1^{\dagger N-\frac{\alpha+1}{2}} a_2^{\dagger \frac{\alpha+1}{2}} \myup{a_1}{N-\frac{\alpha-1}{2}} \myup{a_2}{\frac{\alpha-1}{2}}\rangle.
\label{N_odd_first_coherences}
\end{multline}
In contrast to the case for even $N$, we can arrive at a system of equations similar to Eq.\,(\ref{N_odd_populations}) with only $N$ different terms, which we can solve immediately. Multiplying all equations with $e^{i \phi_k}$ with $2 \leq k \leq (N+1)/2$ gives all other necessary equations in a form equivalent to Eq.\,(\ref{N_even_coherences}):
\begin{widetext}
\begin{multline}
\frac{e^{i \phi_k}}{N+2} x_{\theta_j} = \sum_{\alpha \in G_\beta} \binom{N}{\frac{\alpha+\beta}{2}} \binom{N}{\frac{\alpha-\beta}{2}} (\cos \theta_j)^{2N-\alpha} (\sin \theta_j)^\alpha \langle a_1^{\dagger N-\frac{\alpha-\beta}{2}} a_2^{\dagger \frac{\alpha-\beta}{2}} \myup{a_1}{N-\frac{\alpha+\beta}{2}} \myup{a_2}{\frac{\alpha+\beta}{2}}\rangle\\
+ \sum_{\alpha' \in G_{\beta'}} \binom{N}{\frac{\alpha'+\beta'}{2}} \binom{N}{\frac{\alpha'-\beta'}{2}} (\cos \theta_j)^{2N-\alpha'} (\sin \theta_j)^{\alpha'} \langle a_1^{\dagger N-\frac{\alpha'-\beta'}{2}} a_2^{\dagger \frac{\alpha'-\beta'}{2}} \myup{a_1}{N-\frac{\alpha'+\beta'}{2}} \myup{a_2}{\frac{\alpha'+\beta'}{2}}\rangle,
\label{N_odd_higher_coherences}
\end{multline}
\end{widetext}
again with $\beta = k$, but $\beta' = N-k+2$. Since $N$ is odd, all terms now depend on a different power of $N$, so that the system of linear equations again corresponds to a solvable $(N+1)\times (N+1)$ matrix, making it possible to determine all remaining coherences. For $N=1$ (i.e., simply a polarization measurement), this leads to the choice of measuring the averages
\begin{align}
\begin{array}{l}
\langle a_1^\dagger a_1 \rangle,\\
\langle a_1^\dagger a_1\rangle + \langle a_2^\dagger a_2 \rangle + e^{i \frac{2\pi}{3}} \langle a_1^\dagger a_2 \rangle + e^{-i \frac{2\pi}{3}} \langle a_2^\dagger a_1 \rangle,\\
\langle a_1^\dagger a_1\rangle + \langle a_2^\dagger a_2 \rangle + e^{i \frac{4\pi}{3}} \langle a_1^\dagger a_2 \rangle + e^{-i \frac{4\pi}{3}} \langle a_2^\dagger a_1 \rangle,\\
\langle a_1^\dagger a_1\rangle + \langle a_2^\dagger a_2 \rangle + \langle a_1^\dagger a_2 \rangle + \langle a_2^\dagger a_1 \rangle,
\end{array}
\end{align}
which is different from what is discussed in standard textbooks~\cite{Born:1999}, but equally optimal.

\section{Example and more general applications}
\label{sec_example}
In this section, we discuss the simplest nontrivial example, namely measuring all second-order coherences (case $N = 2$). For this task, we start by using the notation of Simon and Makunda~\cite{Simon:1990} and write the unitary transformation of Eq.~(\ref{unitary_transformation}) in terms of three Euler angles $\xi,\eta,\zeta$:
\begin{align}
U(\xi,\eta,\zeta) = \exp \left(-i\frac{\xi \sigma_2}{2} \right)\, \exp \left(i\frac{\eta \sigma_3}{2} \right) \, \exp \left(-i\frac{\zeta \sigma_2}{2} \right),
\end{align}
where $\sigma_2$ and $\sigma_3$ are Pauli matrices. Simon and Makunda show that the relation of the Euler angles to the actual angles of the three birefringent plates is then given by~\cite{Simon:1990}
\begin{subequations}
\begin{align}
\alpha_{\text{QP}_1} &= \frac{\xi}{2} + \frac{\pi}{4},\\
\alpha_{\text{QP}_2} &=\frac{\xi + \eta}{2} + \frac{\pi}{4},\\
\alpha_{\text{HP}} &=\frac{\xi + \eta - \zeta}{4} -\frac{\pi}{4},
\end{align}
\label{angles}
\end{subequations}
with the Euler angles $\xi$, $\eta,$ and $\zeta$ a function of the abstract angles $\theta$ and $\phi$:
\begin{align}
\cos \frac{\eta}{2} &= \sqrt{a^2 + c^2},\nonumber\\
\exp \left(i \frac{\xi + \zeta}{2}\right) &= \frac{c - ia}{\sqrt{a^2+c^2}},\nonumber\\
\exp \left(i \frac{\xi - \zeta}{2}\right) &= \frac{i b}{|b|},\nonumber
\end{align}
where $a = \text{Re}(e^{i \phi})\sin \theta$, $b = \text{Im}(e^{i \phi})\sin \theta$, and $c = \cos \theta$. In the case that $a=c=0$, which occurs for $\theta, \phi = \pm \frac{\pi}{2}$, the corresponding Euler angles may be chosen as $\eta = \xi = 0$ and $\zeta = 2 \phi$, while in case that $b = 0$, which occurs for $\phi = 0,\pi$, the corresponding Euler angles may be chosen as $\eta = \xi = 0$ and $\zeta = -2 \theta$ $(\zeta =2\theta)$ for $\phi = 0$ $(\phi = \pi)$.

With this translation of two abstract parameters into experimental quantities, it is now straightforward to calculate the settings for our wave plates for any arbitrary measurement. For example, if we wish to measure the nine variances of the Stokes parameters [Eq.\,(\ref{stokes_variances})], the recipe from the previous section tells us to measure the second-order intensity after application of the nine unitary transformations that arise from all possible combinations of $\theta = \frac{1}{8}\pi, \frac{1}{4}\pi, \frac{3}{8}\pi$ and $\phi = \frac{2}{3}\pi, \frac{4}{3}\pi,0$. According to Eqs.\,(\ref{angles}), every pair $(\theta,\phi)$ corresponds to a certain triple of Euler angles $(\xi, \eta, \zeta)$ and this in turn to a certain triple of angles for the wave plates (QP$_1$,QP$_2$,HP), all of which are given in Table\,\ref{table}.
\begin{table}
\renewcommand{\arraystretch}{1.3}
\begin{tabular*}{26em}{@{\extracolsep{\fill}}|c|c|c|}
\hline
$(\theta,\phi)$ & Euler angles $(\xi, \eta, \zeta)$ & ($\alpha_{\text{QP}_1}$,$\alpha_{\text{QP}_2}$,$\alpha_{\text{HP}})$\\
\hline
\hline
$(\frac{1}{8}\pi,\frac{2}{3}\pi)$ & $(1.775,0.676,4.197)$ & $(1.673,2.011,0.169)$\\
$(\frac{1}{4}\pi,\frac{2}{3}\pi)$ & $(2.034,1.318,5.176)$ & $(1.802,2.461,0.329)$\\
$(\frac{3}{8}\pi,\frac{2}{3}\pi)$ & $(-3.833,1.855,-0.692)$ & $(2.010,2.938,0.464)$\\
$(\frac{1}{8}\pi,\frac{4}{3}\pi)$ & $(4.917,0.676,1.775)$ & $(0.102,0.440,1.740)$\\
$(\frac{1}{4}\pi,\frac{4}{3}\pi)$ & $(-1.107,1.318,-4.249)$  & $(0.232,0.891,1.900)$\\
$(\frac{3}{8}\pi,\frac{4}{3}\pi)$ & $(5.591,1.855,2.450)$  & $(0.439,1.367,2.034)$\\
$(\frac{1}{8}\pi,0)$ & $(0,0,-\frac{1}{4}\pi)$ & $(\frac{1}{4}\pi,\frac{1}{4}\pi,\frac{13}{16}\pi)$ \\
$(\frac{1}{4}\pi,0)$ & $(0,0,-\frac{1}{2}\pi)$ & $(\frac{1}{4}\pi,\frac{1}{4}\pi,\frac{7}{8}\pi)$\\
$(\frac{3}{8}\pi,0)$ & $(0,0,-\frac{3}{4}\pi)$ & $(\frac{1}{4}\pi,\frac{1}{4}\pi,\frac{15}{16}\pi)$\\
\hline
\end{tabular*}
\caption{All nine values of the Euler angles $\xi$, $\eta$, and $\zeta$ and the angles of the three wave plates in dependence of the parameters $\theta$ and $\phi$ for measuring second-order coherences.}
\label{table}
\end{table}

From this set of measurements of the second-order intensity moment, it is possible to calculate all variances of the Stokes parameters. In the case of a two-photon Fock state, this corresponds to all density matrix elements (i.e., to a full state tomography). However, our method also serves for the determination of higher-order coherences of other (classical or nonclassical) states. For example, recently the covariance matrix of a Gaussian output state of an optical parametric oscillator has been measured~\cite{DAuria:2009}. If one would want to verify the Gaussian property of this state, the measurement of higher-order coherences like the ones discussed in the present work is required.

The scheme presented here is also not limited to photons of linear polarization: $\hat{a}_1$ and $\hat{a}_2$ may just as well correspond to any other pairwise orthogonal photon polarization modes. In fact, the general idea is applicable to any kind of bosonic multiqubit state where the equivalents to the needed devices exist: a universal SU(2) gadget, a filter which transmits only one of the two qubit states, and a detector capable of performing a correlation measurement on the incident qubits. For example, one possible application could be the characterization of Laguerre-Gaussian beams with photons distributed among two different LG$_{nm}$ modes~\cite{Lassen:2009,Hsu:2009}. In this case, Agarwal discussed the SU(2) structure of their Poincar\'e sphere~\cite{Agarwal:1999}; the equivalent of a polarizing beam splitter can be implemented with holograms~\cite{Mair:2001}, and Ref. \cite{Allen:1992} points at the possibility of constructing an SU(2) gadget consisting of astigmatic lenses.

\section{Conclusion}
In conclusion, it was shown that a very simple experimental setup consisting of two quarter-wave plates, one half-wave plate, a polarizing beam splitter, and a measurement of higher-order intensity moments allows for an optimal measurement of arbitrary-order coherences between two orthogonally polarized modes in a single light beam. Explicit formulas are given for the settings of the three involved wave plates. With these settings, the measurements allow the coherences to be obtained by a solvable system of linear equations. The concept has been exemplified for the case $N=2$, whereby, in the case of a Fock state, the capability of the method to perform a full state tomography has been demonstrated.
The scheme could also be extended to include the measurement of phase-sensitive moments like $\langle a_1 a_2\rangle$; however, in this case, one would need to add a local oscillator before the detector in Fig\,\ref{setup}. In a recent paper \cite{DAuria:2009}, the measurement of such phase-sensitive expectation values was reported for a Gaussian state. For the verification of the Gaussian property of such a state, the measurement of the $N$th-order intensity moments like the ones presented in this paper is required. 
Finally, it was outlined that the method can be fruitfully applied to other systems as the general idea is not limited to linear polarization optics but is applicable to all bosonic systems where a universal SU(2) gagdet and the analogs to a polarizing beam splitter and an intensity moment measurement can be constructed. This property might make the present work interesting for a large range of similar topics.

\section*{Acknowledgment}
U.S. thanks the Elite Network of Bavaria for financial support.

\end{document}